\documentstyle[12pt]{article}
\begin{document}
\def\Wgo {\stackrel{\mbox{\tiny $FNS$ }}{\longrightarrow}}
\def\mbn{\mbox{\boldmath$\nabla$}}
\def\intsum {\int\!\!\!\!\!\!\!{\ss \sum}}
\def\intsums {\int\!\!\!\!\!{\ss\Sigma}}
\def\rbh {\hat{\bf r}}
\def\Lb {{\bf L}}
\def\Xb {{\bf X}}
\def\Yb {{\bf Y}}
\def\Zb {{\bf Z}}
\def\rb {{\bf r}}
\def\kb {{\bf k }}
\def\qb {{\bf q}}
\def\kr {{\bf k}\cdot{\bf r}}
\def\pb {{\bf p}}
\def\etal{{\it et al., }}
\def\cf{{\it cf }}
\def\ie{{\em i.e., }}
\def\etc{ {\it etc}}
\def\dket#1{||#1 \rangle}
\def\dbra#1{\langle #1||}
\def\isim{\:\raisebox{-0.5ex}{$\stackrel{\textstyle.}{=}$}\:}
\def\lsim{\:\raisebox{-0.5ex}{$\stackrel{\textstyle<}{\sim}$}\:}
\def\gsim{\:\raisebox{-0.5ex}{$\stackrel{\textstyle>}{\sim}$}\:}
\def\a {{\alpha}}
\def\b {{\beta}}
\def\e {{\epsilon}}
\def\g {{\gamma}}
\def\r {{\rho}}
\def\s {{\sigma}}
\def\k {{\kappa}}
\def\l {{\lambda}}
\def\m {{\mu}}
\def\n {{\nu}}
\def\r {{\rho}}
\def\t {{\tau}}
\def\w {{\omega}}
\def\be{\begin{equation}}
\def\ee{\end{equation}}
\def\br{\begin{eqnarray}}
\def\er{\end{eqnarray}}
\def\brn{\begin{eqnarray*}}
\def\ern{\end{eqnarray*}}
\def\x{\times}
\def\go{\rightarrow  }
\def\rf#1{{(\ref{#1})}}
\def\nn{\nonumber }
\def\ket#1{|#1 \rangle}
\def\bra#1{\langle #1|}
\def\Ket#1{||#1 \rangle}
\def\Bra#1{\langle #1||}
\def\ov#1#2{\langle #1 | #2  \rangle }
\def\hf {{1\over 2}}
\def\hw {\hbar \omega}
\def\mbs{\mbox{\boldmath$\sigma$}}
\def\gsim{\:\raisebox{-0.5ex}{$\stackrel{\textstyle>}{\sim}$}\:}
\def\lsim{\:\raisebox{-0.5ex}{$\stackrel{\textstyle<}{\sim}$}\:}
\def\mbn{\mbox{\boldmath$\nabla$}}
\def\sss{\scriptscriptstyle}
\def\ss{\scriptstyle}
\def\endauthors{}
\def\authors#1\endauthors{#1}
\def\ex#1{\langle #1 \rangle }
\def\ninj#1#2#3#4#5#6#7#8#9{\left\{\negthinspace\begin{array}{ccc}
#1&#2&#3\\#4&#5&#6\\#7&#8&#9\end{array}\right\}}
\def\sixj#1#2#3#4#5#6{\left\{\negthinspace\begin{array}{ccc}
#1&#2&#3\\#4&#5&#6\end{array}\right\}}
\def\threej#1#2#3#4#5#6{\left(\negthinspace\begin{array}{ccc}
#1&#2&#3\\#4&#5&#6\end{array}\right)}
\def\sixja#1#2#3#4#5#6{\left\{\negthinspace\begin{array}{ccc}
#1&#2&#3\\#4&#5&#6\end{array}\right\}}
\def\ul{\underline}
\def\ol{\overline}
\def\kh {\hat{k}}
\def\rh {\hat{r}}
\def\slash#1{\not\!{#1}}
\def\eabc{\epsilon^{abc}}
\def\rhn {\hat{\rb}_n}
\def\rhm {\hat{\rb}_m}
\def\khb {\hat{\kb}}
\def\rbh{\hat{\rb}}
\begin{titlepage}
\pagestyle{empty}
\baselineskip=21pt
\vskip .2in
\begin{center}
{\large{\bf Nuclear moments for the neutrinoless double beta decay II}}
\end{center}

\vskip .1in
\authors
\centerline{C.~Barbero${}^{\dagger}$, F.~Krmpoti\'{c}${}^{\dagger}$,
A.~Mariano${}^{\dagger}$}
\vskip .15in
\centerline{\it Departamento de F\'\i sica, Facultad de Ciencias}
\centerline{\it
Universidad Nacional de La Plata, C. C. 67, 1900 La Plata, Argentina.}
\vskip .1in
\centerline{and}
\centerline{D.~Tadi\'{c}}
\centerline{\it Physics Department, University of Zagreb}
\centerline{\it Bijeni\v cka c. 32 -10.000- Zagreb, Croatia.}
\endauthors
\endauthors
\vskip 0.5in
\centerline{ {\bf Abstract} }
\baselineskip=18pt
\bigskip
\noindent

The recently developed formalism for the evaluation of
nuclear form factors in neutrinoless double beta decay
is applied to $^{48}Ca$, $^{76}Ge$, $^{82}Se$, $^{100}Mo$, $^{128}Te$
and $^{130}Te$ nuclei.
Explicit  analytical expressions that follows from this theoretical development,
in the single mode model for the decay of $^{48}Ca$, have been worked out.
They are useful both for testing the full numerical calculations, and for
analytically checking the consistency  with other formalisms.
Large configuration space calculations are compared with previous studies,
where alternative formulations were used.
Yet, besides using the G-matrix as residual interaction, we here use a simple
$\delta$-force. Attention is paid to the connected effects of the short
range nuclear correlations and the finite nucleon size.
Constraints on lepton number violating terms in the weak Hamiltonian
(effective neutrino Majorana mass and effective right-handed current
coupling strengths) are deduced.

\vspace{0.5in}

{\it PACS}: 13.15+q;14.80;21.60Jz;23.40

\vspace{0.5in}

$^{\dagger}$Fellow of the CONICET from Argentina.
\end{titlepage}
\baselineskip=18pt

\section{Introduction}

During the last years we have developed a new formulation for the
neutrinoless double beta ($\b\b_{0\n}$) decay, based on the Fourier-Bessel
multipole expansion of the hadronic current, and on the angular momentum
recoupling. First, we did it for the
mass term within the single mode model (SMM) \cite{Krm92}. Later on,
full QRPA calculations  were done for this term \cite{Krm94}. More recently,
the same procedure has been applied to the evaluation of the so called
"recoil term" in the charged Majoron emission \cite{Bar96}. Finally,
the complete formalism, including the right-handed $(V+A)$ hadronic
current, was presented \cite{Bar98}.
The physical substratum in this development is the same
as in the previous works on the same issue
\cite{Hax84,Doi85,Tom91,Ver90,Suh90}, namely,
the same weak Hamiltonian was used. Thus, one
cannot expect to get sensibly different results for the corresponding
observables.
Yet, we have succeeded in
expressing all nuclear $\b\b_{0\n}$ moments in terms of the matrix elements
of {\em only three} well-known one-body spherical tensor operators:
\begin{eqnarray}
{\sf Y}_{\lambda JM}^{\kappa}(k)&=&\sum_n\tau_n^+r_n^\kappa
j_\lambda(kr_n)Y_{JM}(\hat{\bf r}_n),
\nonumber\\
{\sf S}_{\lambda LJM}^{\kappa}(k)&=&\sum_n
\tau_n^+r_n^{\kappa}j_\lambda(kr_n)[\mbs_n\otimes 
Y_{L}(\hat{\bf r}_n)]_{JM},
\label{1}\\
{\sf P}_{LJM}(k)&=&\sum_n\tau_n^+j_L(kr_n)[{\bf p}_n\otimes
Y_{L}(\hat{\bf r}_n)]_{JM},\nonumber
\end{eqnarray}
which have been around in nuclear physics for more than 40 years
\cite{Ros54,Wal66}.
This makes our formulation to be specially suitable for the nuclear
structure calculations, and more simple than other formulations
\cite{Hax84,Doi85,Tom91,Ver90,Suh90}.

In fact, the Fourier-Bessel multipole expansion has also been used by Vergados
\cite{Ver90} and by Suhonen, Khadkikar and Faessler \cite{Suh90}, as the
staring point. However, the final outcomes for the nuclear matrix elements in
these two theoretical developments are quite dissimilar, not only to our
formulas, but also to each other. They are also different to the formulas
derived by Haxton and Stephenson \cite{Hax84} and by Doi, Kotani and Takasugi
\cite{Doi85} within the closure approximation.
As a consequence, the alternative formalisms cannot be confronted analytically,
and the only way to test the consistency among them is by way of numerical
procedures.

As far as we know, numerical calculations, using the formalism
from ref. \cite{Bar98}, have so far been performed only
for the neutrino mass term \cite{Krm94,Sim97}.
Thus, to complete our study on the $\b\b_{0\n}$ matrix elements,
in this paper we carry out calculations for
several $\b\b$ decaying nuclei that are attractive from the experimental
point of view ($^{48}Ca$, $^{76}Ge$, $^{82}Se$, $^{100}Mo$ $^{128}Te$ and
$^{130}Te$). A comparison with similar studies is also made.

As most of the previous studies were performed in the framework
of the QRPA model \cite{Sim97,Tom87,Mut89,Pan96,Fae98}, we will use
here mostly the same nuclear structure approach.
Only the
$^{48}Ca\go ^{48}Ti$-decay will be discussed in a simple shell model,
in order to compare our formalism with that of Vergados \cite{Ver90}.
Also, for the sake of comparison, throughout this work
the bare axial vector
coupling constant $g_{\sss A}=1.254$ will be used, although the effective value
$g_{\sss A}^{\sss eff}=1$ is preferable in nuclear physics \cite{Bro85,Rho97}.

The outline of this paper is as follows:

 In Section 2 we summarize the main
results for the new formalism developed recently \cite{Bar98}.

In Section 3 the decay of $^{48}Ca$ is discussed within the SMM and a
comparison is made with the shell model calculation performed by Pantis
and Vergados \cite{Pan90}.
Here we also derive the analytical expressions for nuclear moments,
which we later use to test the numerical results.
The competition between the effects of the finite nucleon size (FNS)
and the two-nucleon short range correlations (SRC) is discussed as well.

Section 4 deals with full QRPA calculations for $^{48}Ca$, $^{76}Ge$,
$^{82}Se$, $^{100}Mo$, $^{128}Te$ and $^{130}Te$.
We compare them with
two previous QRPA evaluations, namely with: 1) the results obtained
by Muto, Bender
and Klapdor (MBK) \cite{Mut89}, where
the formalism of Doi, Kotani and Takasugi \cite{Doi85} has been
utilized, and 2) the calculation performed by
Pantis, \v Simkovic, Vergados and Faessler (PSVF) \cite{Pan96}
in the framework of the formalism developed by Vergados \cite{Ver90}.
In this section we also show the limits on
the $\b\b_{0\n}$ coupling constants that we deduce from the
most recent experimental data \cite{Mor96,Kla98,Eji96,Ber92,Ale94}.

Concluding remarks are pointed out in Section 5.

\newpage
\section{Neutrinoless double beta decay formalism}

The  $\b\b_{0\n}$ half-life is expressed in the standard form \cite{Tom91}:
\begin{eqnarray}
[T_{0\nu}(0^+\rightarrow  0^+)]^{-1}
&=&\ex{m_\nu}^2C_{mm}+\ex{\lambda}^2C_{\lambda\lambda}+\ex{\eta}^2C_{\eta\eta}\\
&+&\ex{m_\nu}\ex{\lambda}C_{m\lambda}+\ex{m_\nu}\ex{\eta}C_{m\eta}+\ex{\lambda}\ex{\eta}C_{\lambda\eta},
\nonumber
\label{2}\end{eqnarray}
where
$\ex{m_\nu}$ is the effective neutrino mass and
$\ex{\lambda}$ and $\ex{\eta}$ are the
effective coupling constants of the $(V+A)$ hadronic currents.
The coefficients
\begin{eqnarray}
C_{mm}&=&(M_{\sss F}-M_{\sss GT})^2{\cal G}_{1},\nonumber\\
C_{\lambda\lambda}&=&M_{2-}^2{\cal G}_2+\frac{1}{9}M_{1+}^2{\cal G}_4
-\frac{2}{9}M_{2-}M_{1+}{\cal G}_3,\nonumber\\
C_{\eta\eta}&=&M_{2+}^2{\cal G}_2+\frac{1}{9}M_{1-}^2{\cal G}_4
-\frac{2}{9}M_{2+}M_{1-}{\cal G}_3+M_{\sss R}^2{\cal G}_9+M_{\sss R}M_{\sss P}{\cal G}_7
+M_{\sss P}^2{\cal G}_8,\nonumber\\
C_{m\lambda}&=&(M_{\sss F}-M_{\sss GT})\left[M_{2-}{\cal G}_3-M_{1+}{\cal G}_4\right],
\label{3}\\
C_{m\eta}&=&-(M_{\sss F}-M_{\sss GT})\left[M_{2+}{\cal G}_3-M_{1-}{\cal G}_4
+M_{\sss R}{\cal G}_6+M_{\sss P}{\cal G}_5\right],\nonumber\\
C_{\lambda\eta}&=&-2M_{2-}M_{2+}{\cal G}_2+\frac{2}{9}\left[M_{2-}M_{1-}+
M_{2+}M_{1+}\right]{\cal G}_3-\frac{2}{9}M_{1-}M_{1+}{\cal G}_4,
\nonumber\end{eqnarray}
contain the combinations of the matrix elements
\begin{eqnarray}
M_{1\pm}&=&{M}_{\sss GT'}-6M_{\sss T}\pm 3{M}_{\sss F'},\nonumber\\
M_{2\pm}&=&M_{{\sss GT}{\omega}}\pm M_{{\sss F}{\omega}}
-\frac{1}{9}M_{1\mp}.
\label{4}\end{eqnarray}
The kinematical factors
${\cal G}_k$ are given by eq. (3.5.17) in ref. \cite{Doi85} and
the nuclear matrix elements read \cite{Bar98}:
\br
M_{{\sss F}}&=&\left(\frac{g_{{\sss V}}}{g_{{\sss A}}}\right)^2
\sum_{J_\a^\pi pnp'n'}(-)^J{\cal W}_{J0J}(pn){\cal W}_{J0J}(p'n')
{\cal R}^{00}_{JJ}(pnp'n';{\omega}_{J^\pi_{\a}})\rho^{ph}(pnp'n';J_\a^\pi),
\label{5}\er
\br
M_{{\sss GT}}&=&-\sum_{LJ_\a^\pi pnp'n'}(-1)^{L}
{\cal W}_{L1J}(pn){\cal W}_{L1J}(p'n'){\cal R}^{00}_{LL}(pnp'n';{\omega}_{J^\pi_{\a}})
\rho^{ph}(pnp'n';J_\a^\pi),
\label{6}\er
\br
M_{{\sss F'}}&=&-2\left(\frac{g_{{\sss V}}}{g_{{\sss A}}}\right)^2
\sum_{LJ_\a^\pi pnp'n'}i^{L+J+1}(J1|L)(J1|L){\cal W}_{J0J}(pn){\cal W}_{J0J}(p'n') \nn\\
&{\times}& 
{\cal R}^{11}_{JL}(pnp'n';{\omega}_{J^\pi_{\a}})\rho^{ph}(pnp'n';J_\a^\pi),
\label{7}\er
\br
M_{{\sss GT'}}&=&2\sum_{LL'J_\a^\pi pnp'n' }i^{L+L'+1}(L'1|L)(L'1|L)
{\cal W}_{L'1J}(pn){\cal W}_{L'1J}(p'n')\nn\\
&{\times}&
{\cal R}^{11}_{L'L}(pnp'n';{\omega}_{J^\pi_{\a}})\rho^{ph}(pnp'n';J_\a^\pi),
\label{8}\er
\br
M_{{\sss R}}&=&\frac{{\sf R}}{2M_{\sss N}} \frac{g_{{\sss V}}}{g_{{\sss A}}}
\sum_{LL'J_\a^\pi pnp'n'}i^{L+L'}{\cal W}_{L1J}(pn)\rho^{ph}(pnp'n';J_\a^\pi)\nn\\
&\x&
\left\{-f_{\sss W}\left[\frac{}{}\delta_{LL'}-(J1|L)(J1|L')\right]
{\cal W}_{L'1J}(p'n'){\cal R}^{20}_{LL'}(pnp'n';{\omega}_{J^\pi_{\a}})
\right.\nn\\
&-&\left.2\sqrt{6}\hat{L}W(LJ11;1L')(L1|L')
\sum_{\k=\pm}{\cal W}_{L'J}^{(\k)}(p'n'){\cal R}^{1\k}_{LL'}(pnp'n';{\omega}_{J^\pi_{\a}})
\right\},
\label{9}\er
\br
M_{{\sss T}}&=&10\sum_{LL'J'J_\a^\pi pnp'n'}i^{L+L'+1}{\hat{L}^2}(1L|J')(1L|L')
W(12LJ';1L')W(12JJ';1L')\nn\\
&{\times}& {\cal W}_{L'1J}(pn){\cal W}_{J'1J}(p'n')
{\cal R}^{11}_{L'L}(pnp'n';{\omega}_{J^\pi_{\a}})\rho^{ph}(pnp'n';J_\a^\pi),
\label{10}\er
\br
M_{{\sss P}} &=&2\sqrt{6}\frac{g_{{\sss V}}}{g_{{\sss A}}}
\sum_{LJ_\a^\pi pnp'n'}i^{1-L-J}\hat{J}(J1|L)(J1|L)W(JL11;1J)\nn\\
&{\times}&{\cal W}_{J0J}(pn){\cal W}_{J1J}(p'n')
{\cal R}^{11}_{JL}(pnp'n';{\omega}_{J^\pi_{\a}})\rho^{ph}(pnp'n';J_\a^\pi).
\label{11}\er
Here 
${\sf R}$ is the nuclear radius, $M_{\sss N}$ is the nucleon mass,
the index $\a$ labels different intermediate states with
the same spin $J$ and parity $\pi$, $\hat{J}=\sqrt{2J+1}$
and $(L1|J)$ is a short notation for the Clebsh-Gordon coefficient $(L010|J0)$.
The angular momentum coefficients are:
\footnote{We use here the angular momentum coupling
$\ket{({1 \over 2},l)j}$.}
\begin{eqnarray}
{\cal W}_{LSJ}(pn)&=&\sqrt{2}\hat{S} \hat{J}\hat{L}\hat{l}_n\hat{j}_n\hat{j}_p
(l_nL|l_p) \ninj{l_p}{{1 \over 2}}{j_p}{L}{S}{J}{l_n}{{1 \over 2}}{j_n},
\nonumber\\
{\cal W}_{LJ}^{(\pm)}(pn)&=&\mp i(-1)^{j_p+l_n+L+{1 \over 2}} \hat{J}\hat{L}
\hat{l}_p \hat{j}_p\hat{j}_n (l_n+{\ss{1 \over 2}}\mp
{\ss{1 \over 2}})^{{1 \over 2}}(l_pL|l_n\mp 1)\nonumber\\
&{\times}&W(l_pj_pl_nj_n {\ss{1 \over 2}}J)
W(LJl_n\mp 1l_n;1l_p),
\label{12}\end{eqnarray}
and the two-body radial integrals are defined as
\begin{equation}
{\cal R}^{\kappa\kappa'}_{LL'}(pnp'n';{\omega}_{J^\pi_{\a}})=
{\sf R}\int dk k^{2+\kappa}v(k;{\omega}_{J^\pi_{\a}})R_L^0(pn;k) 
R_{L'}^{\kappa'}(p'n';k),~~~~\k'=0,1,\pm,
\label{13}\end{equation}
where
\begin{equation}
v(k;{\omega}_{J^\pi_{\a}}) =\frac{2}{\pi} \frac{1}{k(k+\omega_{J^\pi_{\a}})},
~~~~~~\omega_{J^\pi_{\a}} =E_ {J^\pi_{\a}} -\frac{1}{2}\left(E_i +E_f\right),
\label{14}\end{equation}
is the "neutrino potential", and
\begin{eqnarray}
R^\kappa_L(pn;k) &\equiv&R^\kappa_L(l_p,n_p,l_n,n_n;k)=
\int_0^\infty u_{n_p,l_p}(r)u_{n_n,l_n}(r)j_L(kr)r^{2+\kappa} dr,
\nonumber\\
R_L^{(\pm)}(pn;k)&=&\int_0^\infty u_{n_p,l_p}(r)\left(\frac{d}{dr}\pm
\frac{2l_n+1\pm 1}{2r}\right)u_{n_n,l_n}(r)j_L(qr)r^2 dr.
\label{15}\end{eqnarray}
There are also two additional matrix elements, namely $M_{{\sss F}{\omega}}$
and  $M_{{\sss GT}{\omega}}$, which are obtained from $M_{{\sss F}}$ and
$M_{{\sss GT}}$ by the replacement
\begin{equation}
v(k;{\omega}_{J^\pi_{\a}}) \rightarrow v_{{\omega}} (k;{\omega}_{J^\pi_{\a}})
=\frac{2}{\pi} \frac{1}{(k+ \omega_{J^\pi_{\a}} )^2}.
\label{16}\end{equation}
Finally the two-body state dependent particle-hole (ph) density matrix
\begin{eqnarray}
\rho^{ph}(pnp'n';J_\a^\pi) &=&\hat{J}^{-2} \Bra{0^+_f}
(a^{{\dagger}}_{p}a_{\bar{n}})_{J^\pi} \Ket{{J_\a^\pi}}
\Bra{{J_{\a}^\pi}}(a^{{\dagger}}_{p'} a_{\bar{n}'})_{J^\pi} \Ket{ 0^+_i},
\label{17}\end{eqnarray}
contains information on the wave functions of the initial ($\ket{ 0^+_i}$),
final ($\ket{ 0^+_f}$), and virtual intermediate ($\ket{{J_\a^\pi}}$) states.

In particular, within the QRPA formulation, and after solving both the BCS
and the RPA equations for the intermediate $(N-1,Z+1)$ nucleus \cite{Krm97},
the two-body density matrix becomes
\begin{eqnarray}
{\rho}^{ph}(pnp'n';J_{\a}^\pi)=
\left[u_nv_pX_{J^\pi_{\a}}(pn)+ u_pv_nY_{J_{\a}^\pi}(pn)\right]
\left[ u_{p'}v_{n'}X_{J^\pi_{\a}}(p'n')+u_{n'}v_{p'}Y_{J^\pi_{\a}}(p'n')
\right],
\label{18}\end{eqnarray}
where the notation has the standard meaning \cite{Krm94,Krm97}.
\newpage
\section{$^{48}Ca\go ^{48}Ti$ decay within the
single mode model}

In the SMM the virtual states in the intermediate nucleus $^{48}Sc$ are:
$[0f_{7/2}(p)0f_{7/2}(n)]_{J^+}$, where $J^+=0^+\cdots 7^+$ \cite{Krm94}.
When the harmonic oscillator radial wave functions are
used  and the excitation energy ${\w}_{J^\pi_{\a}}$ is taken to be
zero, we can go a step further in the analytical calculations.

\begin{table}[h]
\begin{center}
\caption {Radial integrals ${\cal R}^{\k\k'}_{LJ}(pnpn;{\w}_{J^\pi_{\a}})$
for the $\b\b_{0\n}$ decay in $^{48}Ca$. The excitation energy
${\w}_{J^\pi_{\a}}$ is taken to be zero, and
harmonic oscillator radial wave functions were employed with the
oscillator parameter $\n=M\w/\hbar$.
Short notation $C=\frac{33600\protect{\sqrt{2\pi}}}{{\sf R}\nu
\protect{\sqrt{\n}}}$ has been used.}
\label{tab1}
\bigskip
\begin{tabular}{crrrrrr}
\hline
$L$&$J$&$C\nu{\cal R}^{00}_{LJ}$&$C\nu{\cal R}^{11}_{JL}$&
$C{\cal R}^{20}_{LJ}$&$C{\cal R}^{1+}_{JL}$&$C{\cal R}^{1-}_{JL}$\\
\hline
$\!0\!$&$\!0\!$&$\!37230\!$&$\!\!$&$\!12870\!$&$\!\!$&$\!\!$\\
$\!1\!$&$\!0\!$&$\!\!$&$\!18615\!$&$\!\!$&$\!9805\!$&$\!-18615\!$\\
$\!2\!$&$\!0\!$&$\!\!$&$\!\!$&$\!-690\!$&$\!\!$&$\!\!$\\
$\!0\!$&$\!2\!$&$\!\!$&$\!\!$&$\!-690\!$&$\!\!$&$\!\!$\\
$\!1\!$&$\!2\!$&$\!\!$&$\!11835\!$&$\!\!$&$\!16585\!$&$\!-11835\!$\\
$\!2\!$&$\!2\!$&$\!4734\!$&$\!\!$&$\!12870\!$&$\!\!$&$\!\!$\\
$\!3\!$&$\!2\!$&$\!\!$&$\!11835\!$&$\!\!$&$\!7065\!$&$\!-11835\!$\\
$\!4\!$&$\!2\!$&$\!\!$&$\!\!$&$\!6030\!$&$\!\!$&$\!\!$\\
$\!2\!$&$\!4\!$&$\!\!$&$\!\!$&$\!6030\!$&$\!\!$&$\!\!$\\
$\!3\!$&$\!4\!$&$\!\!$&$\!8415\!$&$\!\!$&$\!10485\!$&$\!-8415\!$\\
$\!4\!$&$\!4\!$&$\!1870\!$&$\!\!$&$\!12870\!$&$\!\!$&$\!\!$\\
$\!5\!$&$\!4\!$&$\!\!$&$\!8415\!$&$\!\!$&$\!5445\!$&$\!-8415\!$\\
$\!6\!$&$\!4\!$&$\!\!$&$\!\!$&$\!8910\!$&$\!\!$&$\!\!$\\
$\!4\!$&$\!6\!$&$\!\!$&$\!\!$&$\!8910\!$&$\!\!$&$\!\!$\\
$\!5\!$&$\!6\!$&$\!\!$&$\!6435\!$&$\!\!$&$\!7425\!$&$\!-6435\!$\\
$\!6\!$&$\!6\!$&$\!990\!$&$\!\!$&$\!12870\!$&$\!\!$&$\!\!$\\
$\!7\!$&$\!6\!$&$\!\!$&$\!6435\!$&$\!\!$&$\!\!$&$\!-6435\!$\\
\hline \end{tabular} \end{center}
\end{table}
The results for the radial integrals ${\cal R}^{\kappa\kappa'}_{LL'}(pnpn;
{\omega}_{J^\pi_{\a}})$ are listed in Table \ref{tab1}, and
after performing the summations on $L,L'$ and $J'$, as indicated in
eqs. \rf{5}-\rf{11}, we get:
\newpage
\br
M_{\sss F}&=&M_{\sss F'}=M_{{\sss F}{\w}}= {\sf R}\sqrt{\frac{\n}{2\pi}}
\left(\frac{g_{{\sss V}}}{g_{{\sss A}}}\right)^2\sum_{J^+}
A_{\sss F}(J^+){\rho}^{ph}(J^+),\nn\\
M_{\sss GT}&=&M_{\sss GT'}=M_{{\sss GT}{\w}}= {\sf R}\sqrt{\frac{\n}{2\pi}}
\sum_{J^+}A_{\sss GT}(J^+){\rho}^{ph}(J^+),\nn\\
M_{\sss R}&=&\frac{{\sf R}^2}{2M_{\sss N}}\sqrt{\frac{\n^3}{2\pi}}
\frac{g_{{\sss V}}}{g_{{\sss A}}}\sum_{J^+}
\left[f_{\sss W}A_{1\sss R}(J^+)+A_{2\sss R}(J^+)\right]{\rho}^{ph}(J^+),
\label{19}\\
M_{\sss T}&=& {\sf R}\sqrt{\frac{\n}{2\pi}}\sum_{J^+}A_{\sss T}(J^+)
{\rho}^{ph}(J^+),~~~~~ M_{{\sss P}}=0,
\nn\er
where $ {\rho}^{ph}(J^+)\equiv {\rho}^{ph}(pnpn;J^+)$.
The coefficients $A_{\sss X}(J^+)$ are given in Table \ref{tab2}.
It should be noted that: i) in the SMM the matrix element
 $M_{{\sss P}}$ is always null,
independently of the value for the excitation energy ${\w}_{J^\pi_{\a}}$, and
ii) while the Fermi matrix elements arise only from even multipoles, the
remaining matrix elements only come from odd multipoles.

\begin{table}[h]
\begin{center}
\caption {Coefficients $A_{{\sss X}}(J^+)$
for the matrix elements given by eq. \protect \rf{19}.}
\label{tab2}
\bigskip
\begin{tabular}{crrrrr}
\hline
$J^+$&$980A_{\sss F}(J^+)$&$63700A_{\sss GT}(J^+)$&$245A_{1\sss R}(J^+)$
&$245A_{2\sss R}(J^+)$&$47775A_{\sss T}(J^+)$\\
\hline
$\!0^+\!$&$\!8687\!$&$\!0\!$&$\!0\!$&$\!0\!$&$\!0\!$\\
$\!1^+\!$&$\!0\!$&$\!-746499\!$&$\!-738\!$&$\!-1218\!$&$\!54327\!$\\
$\!2^+\!$&$\!1315\!$&$\!0\!$&$\!0\!$&$\!0\!$&$\!0\!$\\
$\!3^+\!$&$\!0\!$&$\!-117247\!$&$\!-552\!$&$\!-720\!$&$\!36751\!$\\
$\!4^+\!$&$\!459\!$&$\!0\!$&$\!0\!$&$\!0\!$&$\!0\!$\\
$\!5^+\!$&$\!0\!$&$\!-72575\!$&$\!-870\!$&$\!-510\!$&$\!21275\!$\\
$\!6^+\!$&$\!175\!$&$\!0\!$&$\!0\!$&$\!0\!$&$\!0\!$\\
$\!7^+\!$&$\!0\!$&$\!-91875\!$&$\!-2450\!$&$\!0\!$&$\!18375\!$\\
\hline \end{tabular} \end{center}
\end{table}

In the shell model (SM) the virtual states $\ket{J^+}$ in the $^{48}Sc$
nucleus and the wave function $\ket{0^+_f}$ in the final nucleus $^{48}Ti$,
are described, respectively, as the one-particle one-hole and two-particle
two-hole excitations on the ground state $\ket{0^+_i}$ in $^{48}Ca$. That is:
\br
\ket{J^+}&=& \ket{[f_{7/2}(p)f^{-1}_{7/2}(n)]J^+},~~~~J^+=0^+,1^+\cdots 7^+,
\nn\\
\ket{0^+_f}&=&\sum_{I^+}\ov{I^+}{0^+_f}\ket{
[f^2_{7/2}(p)]I^+,[f^{-2}_{7/2}(n)]I^+;0^+},~~~ I^+=0^+,2^+,4^+,6^+,
\label{20}\er
and the two-body density reads \cite{Bar98}
\footnote{
It is worth remembering \cite{Krm94} that, in the SMM, eq. \rf{18} reduces to
\[ \rho^{ph}(J^+)= \rho^{ph}_{BCS}(J^+)
\left[\frac{\omega_0+G(J^+)} {\omega_J^+}\right], \]
where $\rho^{ph}_{BCS}(J^+)=u_pv_nu_nv_p$ is the BCS value for the two-body
density, $\w_0$ is the unperturbed proton-neutron quasiparticle energy,
$\omega_J^+$ are the QRPA energies, and $G(J^+)\equiv G(pnpn;J^+)$ is the
particle-particle matrix element. Thus we see that, within the SMM,
the last factor in this equation plays the role of the effective charge for the
$\b\b_{0\n}$ decay, induced by the QRPA correlations.}

\be
{\rho}^{ph}(J^+)=-2\sum_{I}\hat{I}\ov{0_f^+}{I^+}(-)^{J}
\sixj{7/2} {7/2}{J}{7/2} {7/2}{I}.
\label{21}\ee

\begin{table}[h]
\begin{center}
\caption { Amplitudes $\ov{0_f^+}{J^+}$ of the ground state wave function in
$^{48}Ti$ and the corresponding coefficients ${\rho}^{ph}(J^+)$.}
\label{tab3}
\bigskip
\begin{tabular}{crr}
\hline
$J^+$& $\ov{0_f^+}{J^+}$& ${\rho}^{ph}(J^+)$\\
\hline
$\!0^+\!$&$\!0.9433\!$&$\!-0.0005\!$\\
$\!1^+\!$&$\!-\!$&$\!0.0426\!$\\
$\!2^+\!$&$\!-0.3126\!$&$\!0.1159\!$\\
$\!3^+\!$&$\!-\!$&$\!0.2098\!$\\
$\!4^+\!$&$\!-0.1092\!$&$\!0.3072\!$\\
$\!5^+\!$&$\!-\!$&$\!0.3533\!$\\
$\!6^+\!$&$\!0.0231\!$&$\!0.3017\!$\\
$\!7^+\!$&$\!-\!$&$\!0.1186\!$\\
\hline \end{tabular} \end{center}
\end{table}

To get close to the results obtained by Pantis and
Vergados \cite{Pan90} as much as possible,
we use the same wave function for the state $\ket{0^+_f}$ that was
employed in their work. This wave function is listed in Table \ref{tab3}
together with the resulting values for the
densities ${\rho}^{ph}(J^+)$. Note that ${\rho}^{ph}(0^+)$ and
${\rho}^{ph}(1^+)$
are relatively small because of the restoration of isospin and
SU(4) symmetries, respectively. Their single-particle (Hartree-Fock) values,
obtained from $\ov{0_f^+}{0^+}=1$, are far larger than the values shown in
Table \ref{tab3} (${\rho}^{ph}(0^+)= -{\rho}^{ph}(1^+)=-0.25$).
On the other hand, they are identically null when
these symmetries are totally restored.

The short-range correlations (SRC) between the two nucleons are taken into
account via the correlation function \cite{Bro77}
\be
f_{SRC}(r)=1-j_0(k_c r),
\label{22}\ee
where $k_c=3.93$ fm$^{-1}$ is roughly the Compton wavelength of the $\w$-meson.
The finite nucleon size (FNS) effects are introduced in the usual way, \ie
by the dipole form factors in momentum space:
\br
(g_{\sss V,A})_{FNS}=g_{\sss V,A}\left(
\frac{\Lambda^2}{\Lambda^2+k^2}\right)^2,
\label{23}\er
with $\Lambda=850$ MeV. The corresponding modifications of the neutrino
potentials are shown in refs. \cite{Krm92,Bar98}.

\begin{table}[h]
\begin{center}
\caption {Nuclear matrix elements for the decay
$^{48}Ca\go ^{48}Ti$ within the single-mode shell-model calculations.
We have used
${\w}_{J^\pi_{\a}}=0$ and four different results are presented:
1) ({\it{bare}}) no correlations and no nucleon form factor,
2) ({\it{FNS}}) no correlations but with nucleon form factor,
3) ({\it{SRC}}) short range correlations but without nucleon form factor,
and 4) ({\it{FNS+SRC}}) correlations and nucleon form factor.}
\label{tab4}
\bigskip
\begin{tabular}{cccccccccc}
\hline
&&$M_{\sss GT}$&$M_{\sss F}$&$M_{{\sss GT}\w}$
&$M_{{\sss F}\w}$&$M_{{\sss GT'}}$&$M_{{\sss F'}}$
&$M_{\sss R}$&$M_{\sss T}$\\
\hline
&\ul {Present Results} &&&&&&&&\\
&bare&$\!-1.168\!$&$\!0.177\!$&$\!-1.168\!$&$\!0.177\!$&$\!-1.168\!$&$\!0.177\!$& $\!-1.435\!$&$\!0.330\!$\\
&SRC&$\!-1.080\!$&$\!0.159\!$&$\!-1.080\!$&$\!0.159\!$&$\!-0.657\!$&$\!0.073\!$&$\!-0.105\!$&$\!0.284\!$\\
&FNS&$\!-0.960\!$&$\!0.134\!$&$\!-0.960\!$&$\!0.134\!$&$\!-0.644\!$&$\!0.066\!$&$\!-0.929\!$&$\!0.312\!$\\
&FNS+SRC&$\!-0.947\!$&$\!0.130\!$&$\!-0.947\!$&$\!0.130\!$&$\!-0.574\!$&$\!0.052\!$&$\!-0.796\!$&$\!0.309\!$\\
\hline
&\ul {Pantis \& Vergados \cite{Pan90}} &&&&&&&&\\
&bare&$\!-1.216\!$&$\!0.185\!$&$\!-1.216\!$&$\!0.185\!$&$\!-1.216\!$&$\!0.185\!$ &$\!-2.178\!$&$\!0.344\!$\\
&SRC&$\!-0.859\!$&$\!0.108\!$&$\!-0.856\!$&$\!0.108\!$&$\!-0.841\!$&$\!0.105\!$ &$\!-0.115\!$&$\!0.346\!$\\
&FNS&$\!-0.986\!$&$\!0.134\!$&$\!-0.986\!$&$\!0.136\!$&$\!-0.635\!$&$\!0.063\!$ &$\!-1.344\!$&$\!0.322\!$\\
&FNS+SRC&$\!-0.731\!$&$\!0.117\!$&$\!-0.731\!$&$\!0.098\!$&$\!-0.532\!$&$\!0.055\!$ &$\!-0.324\!$&$\!0.330\!$\\
\hline
\hline\end{tabular} \end{center}
\end{table}

Present results are confronted with those obtained by
Pantis and Vergados \cite{Pan90} in Table \ref{tab4}.
They used a somewhat different approximation for the SRC and therefore
it is plausible that our matrix elements do not fully agree with 
theirs
in the second and fourth case. In the other two cases, they should be
identical, but they are not! The difference is particularly pronounced
for the recoil matrix elements
$M_{\sss R}$.
The reason for the discrepancies could be the values used for
the harmonic oscillator
parameter $\n=M\w/\hbar$ and the nuclear radius $\sf R$; we have utilized
$\n=0.916 A^{-1/3}$ fm$^{-2}$ and ${\sf R}=1.2 A^{1/3}$ fm.
\footnote{Our definitions for the nuclear matrix elements
$M_{\sss F}, M_{\sss F'}, M_{{\sss F}{\w}}$ and
$M_{\sss R}$
agree with
those of
Pantis and Vergados \cite{Pan90} only for $g_{\sss A}=g_{\sss V}$.
As we used here
$g_{\sss A}=1.254 $, the results listed in their Table 1 have been
renormalized accordingly.}

Anyhow it is worth noting that in both calculations
the FNS effects and the SRC act coherently on the Fermi (F) and
Gamow-Teller (GT) moments, in the
sense that their combined effects always diminish them more than
when they are acting individually. This, however, does not happen with
$M_{\sss R}$, in which case the FNS+SRC values turn out to be significantly
larger than the SRC ones. The explanation for this somewhat curious
behavior of the recoil matrix element was given by Tomoda \etal \cite
{Tom91,Tom86} and is as follows.
The contribution of the weak magnetism in
\rf{9} can be decomposed into the central and tensor parts \cite{Tom91}.
The central part is the dominant one, and within the closure approximation and
for ${\w}_{J^\pi_{\a}}=0$, it can be rewritten in the form:
\footnote{The following relation has been used:
\brn
\delta(\rb-\rb')=
&=&\frac{2}{\pi}\sum_{lm}Y_{lm}(\rh) Y^*_{lm}(\rh')
\int k^2dk j_l(kr)j_{l}(kr').
\ern}
\begin{eqnarray}
M_{{\sss RC}}(bare)=-\frac{4\pi{\sf R}^2}{3M_{\sss N}}\frac{f_{\sss W}
g_{\sss V}}{g_{\sss A}}
\bra{{\ss F}}\sum_{mn}\t^+_m\t^+_n
\mbs_m\cdot\mbs_n \delta(\rb_m-\rb_n)\ket{{\ss I}},
\label{24} \end{eqnarray}
This matrix element is totally killed by the SRC \rf{22} and therefore
\begin{eqnarray}
M_{{\sss RC}}(SRC)=-\frac{4\pi{\sf R}^2}{3M_{\sss N}}\frac{f_{\sss W}
g_{\sss V}}{g_{\sss A}}
\bra{{\ss F}}\sum_{mn}\t^+_m\t^+_n
\mbs_m\cdot\mbs_n \delta(\rb_m-\rb_n)
f_{SRC} (\rb_m-\rb_n) \ket{{\ss I}}\equiv 0.
\label{25} \end{eqnarray}
The $k^2$ dependence of the form factors \rf{23} distributes the $\delta$-function
over a finite region \cite{Tom91,Bro77}, \ie
\be
\delta(\rb)\Wgo \frac{\Lambda^3}{64\pi}e^{-\Lambda r}
\left[1-\Lambda r +\frac{1}{3}(\Lambda r)^2\right].
\label{26}\ee
Consequently, the matrix element \rf{24} decreases
($M_{{\sss RC}}(FNS)\le M_{{\sss RC}}(bare)$) and
$M_{{\sss RC}}(FNS+SRC)\ne 0$.
\newpage
\section {QRPA calculations}
We have employed a residual
$\delta$-force $V=-4\pi(v_sP_s+v_tP_t)\delta(r)$, with different strength
constants $v_s$ and $v_t$ for the particle-hole,  particle-particle and
pairing channels
\cite{Krm94,Krm97,Hir90}.
The single-particle energies, as well as the pairing parameters
$v_s^{pair}(p)$ and $v_s^{pair}(n)$, have been fixed by fitting the
experimental pairing gaps to a Wood-Saxon potential well.

\begin{table}[h]
\begin{center}
\caption {QRPA results for the nuclear matrix elements that include
both the FNS and SRC effects.
An average excitation energy $\ex{\w_{J^\pi_\a}}$ of $5.0$ MeV has
been used in the present evaluation.}
\label{tab5}
\bigskip
\begin{tabular}{ccccccccccc}
\hline
Nucleus&&$M_{\sss GT}$&$M_{\sss F}$&$M_{{\sss GT}\w}$
&$M_{{\sss F}\w}$&$M_{{\sss GT'}}$&$M_{{\sss F'}}$
&$M_{\sss R}$&$M_{\sss T}$&$M_{\sss P}$\\
\hline
$^{48}Ca$&&&&&&&&&\\
&present&$\!-0.953\!$&$\!0.376\!$&$\!-1.010\!$&$\!0.361\!$&$\!-0.022\!$&$\!0.203\!$&$\!-1.888\!$&$\!-0.033\!$&$\!0.075\!$\\
&ref. \cite{Pan96}&$\!-0.785\!$&$\!0.367\!$&$\!-0.830\!$&$\!0.343\!$&$\!-0.765\!$&$\!0.395\!$&$\!-1.522\!$&$\!0.166\!$&$\!-0.131\!$\\
\hline
$^{76}Ge$&&&&&&&&&\\
&present&$\!-2.845\!$&$\!0.749\!$&$\!-2.864\!$&$\!0.723\!$&$\!-0.837\!$&$\!0.371\!$&$\!-4.863\!$&$\!-0.065\!$&$\!-0.889\!$\\
&ref. \cite{Mut89}&$\!-3.014\!$&$\!1.173\!$&$\!-2.912\!$&$\!1.025\!$&$\!-1.945\!$&$\!1.058\!$&$\!-3.594\!$&$\!0.612\!$&$\!0.530\!$\\
&ref. \cite{Pan96}&$\!-2.929\!$&$\!0.111\!$&$\!-2.683\!$&$\!0.111\!$&$\!-3.154\!$&$\!0.102\!$&$\!-7.423\!$&$\!0.714\!$&$\!-3.360\!$\\
\hline
$^{82}Se$&&&&&&&&&\\
&present&$\!-2.717\!$&$\!0.800\!$&$\!-2.769\!$&$\!0.771\!$&$\!-0.603\!$&$\!0.398\!$&$\!-5.147\!$&$\!-0.061\!$&$\!-0.754\!$\\
&ref. \cite{Mut89}&$\!-2.847\!$&$\!1.071\!$&$\!-2.744\!$&$\!0.939\!$&$\!-1.886\!$&$\!0.966\!$&$\!-3.343\!$&$\!0.789\!$&$\!0.500\!$\\
&ref. \cite{Pan96}&$\!-2.212\!$&$\!0.018\!$&$\!-2.124\!$&$\!0.029\!$&$\!-2.323\!$&$\!0.009\!$&$\!-3.700\!$&$\!-0.175\!$&$\!0.108\!$\\
\hline
$^{100}Mo$&&&&&&&&\\
&present&$\!-2.155\!$&$\!0.972\!$&$\!-2.363\!$&$\!0.935\!$&$\!0.354\!$&$\!0.493\!$&$\!-6.150\!$&$\!-0.233\!$&$\!1.265\!$\\
&ref. \cite{Mut89}&$\!-0.763\!$&$\!1.356\!$&$\!-1.330\!$&$\!1.218\!$&$\!1.145\!$&$\!1.161\!$&$\!-4.528\!$&$\!0.823\!$&$\!-1.182\!$\\
&ref. \cite{Pan96}&$\!-0.615\!$&$\!0.471\!$&$\!-0.420\!$&$\!0.436\!$&$\!-0.722\!$&$\!0.512\!$&$\!-0.930\!$&$\!0.293\!$&$\!2.393\!$\\
\hline
$^{128}Te$&&&&&&&&&\\
&present&$\!-3.417\!$&$\!1.019\!$&$\!-3.476\!$&$\!0.980\!$&$\!-0.835\!$&$\!0.451\!$&$\!-6.354\!$&$\!-0.136\!$&$\!-0.560\!$\\
&ref. \cite{Mut89}&$\!-3.103\!$&$\!1.184\!$&$\!-3.011\!$&$\!1.047\!$&$\!-1.999\!$&$\!1.054\!$&$\!-4.371\!$&$\!0.583\!$&$\!0.483\!$\\
&ref. \cite{Pan96}&$\!-2.437\!$&$\!0.044\!$&$\!-2.179\!$&$\!0.029\!$&$\!-2.673\!$&$\!0.054\!$&$\!-1.522\!$&$\!0.748\!$&$\!-3.412\!$\\
\hline
$^{130}Te$&&&&&&&&&\\
&present&$\!-3.225\!$&$\!0.978\!$&$\!-3.271\!$&$\!0.938\!$&$\!-0.819\!$&$\!0.448\!$&$\!-5.934\!$&$\!-0.118\!$&$\!-0.560\!$\\
&ref. \cite{Mut89}&$\!-2.493\!$&$\!0.977\!$&$\!-2.442\!$&$\!0.867\!$&$\!-1.526\!$&$\!0.860\!$&$\!-3.736\!$&$\!0.574\!$&$\!0.387\!$\\
&ref. \cite{Pan96}&$\!-2.327\!$&$\!0.009\!$&$\!-2.083\!$&$\!-0.002\!$&$\!-2.553\!$&$\!0.016\!$&$\!-5.445\!$&$\!0.656\!$&$\!-3.376\!$\\
\hline\end{tabular} \end{center}
\end{table}
\newpage
As already mentioned, the proton and neutron gap equations have been solved
for the intermediate $(N-1,Z+1)$ nucleus as in ref. \cite{Krm97}, and
we deal only with one QRPA equation.
Note that in this procedure we avoid the problem of overlapping
of two sets of the same intermediate states generated from initial and
final nuclei \cite{Fae98}.
\begin{table}[h]
\begin{center}
\caption {The coefficients that appearing in eq. \protect{\rf{2}} (in units of
yr$^{-1}$), evaluated with the matrix elements given in Table \ref{tab5}.
We have used the
kinematical factors from ref. \cite{Doi85}.}
\label{tab6}
\bigskip
\begin{tabular}{cccccccc}
\hline
Nucleus&&$C_{mm}$&$C_{\lambda\lambda}$&$C_{\eta\eta}$&$C_{m\lambda}$&$C_{m\eta}$&$C_{\lambda\eta}$\\
\hline
$^{48}Ca$&&&&&&\\
&present&$\!1.13~10^{-13}\!$&$\!7.42~10^{-13}\!$&$\!1.47~10^{-8}\!$&$\!-9.94~10^{-14}\!$&$\!2.56~10^{-11}\!$&$\!-8.10~10^{-13}\!$\\
&ref. \cite{Pan96}&$\!1.07~10^{-13}\!$&$\!3.68~10^{-13}\!$&$\!6.63~10^{-10}\!$&$\!-4.75~10^{-14}\!$&$\!-5.20~10^{-12}\!$&$\!-3.43~10^{-13}\!$\\
\hline
$^{76}Ge$&&&&&&\\
&present&$\!8.27~10^{-14}\!$&$\!1.26~10^{-13}\!$&$\!8.20~10^{-9}\!$&$\!-4.61~10^{-14}\!$&$\!2.56~10^{-11}\!$&$\!-1.57~10^{-13}\!$\\
&ref. \cite{Mut89}&$\!1.12~10^{-13}\!$&$\!1.36~10^{-13}\!$&$\!4.44~10^{-9}\!$&$\!-4.11~10^{-14}\!$&$\!2.19~10^{-11}\!$&$\!-4.99~10^{-14}\!$\\
&ref. \cite{Pan96}&$\!7.33~10^{-14}\!$&$\!1.12~10^{-13}\!$&$\!3.22~10^{-9}\!$&$\!-4.49~10^{-14}\!$&$\!-1.54~10^{-11}\!$&$\!-2.11~10^{-13}\!$\\
\hline
$^{82}Se$&&&&&&\\
&present&$\!3.48~10^{-13}\!$&$\!1.14~10^{-12}\!$&$\!3.65~10^{-8}\!$&$\!-2.47~10^{-13}\!$&$\!8.83~10^{-11}\!$&$\!-1.39~10^{-12}\!$\\
&ref. \cite{Mut89}&$\!4.33~10^{-13}\!$&$\!1.01~10^{-12}\!$&$\!1.54~10^{-8}\!$&$\!-1.60~10^{-13}\!$&$\!6.37~10^{-11}\!$&$\!-3.84~10^{-13}\!$\\
&ref. \cite{Pan96}&$\!1.75~10^{-13}\!$&$\!4.78~10^{-13}\!$&$\!1.53~10^{-9}\!$&$\!-8.77~10^{-14}\!$&$\!-1.31~10^{-11}\!$&$\!-9.32~10^{-13}\!$\\
\hline
$^{100}Mo$&&&&&&\\
&present&$\!4.47~10^{-13}\!$&$\!1.93~10^{-12}\!$&$\!6.58~10^{-8}\!$&$\!-4.22~10^{-13}\!$&$\!1.32~10^{-10}\!$&$\!-2.03~10^{-12}\!$\\
&ref. \cite{Mut89}&$\!2.05~10^{-13}\!$&$\!1.05~10^{-12}\!$&$\!3.50~10^{-8}\!$&$\!-1.61~10^{-13}\!$&$\!6.48~10^{-11}\!$&$\!7.03~10^{-13}\!$\\
&ref. \cite{Pan96}&$\!6.77~10^{-14}\!$&$\!3.28~10^{-14}\!$&$\!2.91~10^{-9}\!$&$\!-5.64~10^{-15}\!$&$\!-1.11~10^{-11}\!$&$\!2.45~10^{-14}\!$\\
\hline
$^{128}Te$&&&&&&\\
&present&$\!3.60~10^{-14}\!$&$\!1.12~10^{-14}\!$&$\!3.14~10^{-9}\!$&$\!-1.10~10^{-14}\!$&$\!1.42~10^{-11}\!$&$\!-1.25~10^{-14}\!$\\
&ref. \cite{Mut89}&$\!3.36~10^{-14}\!$&$\!7.39~10^{-15}\!$&$\!1.50~10^{-9}\!$&$\!-4.86~10^{-15}\!$&$\!9.46~10^{-12}\!$&$\!-1.87~10^{-15}\!$\\
&ref. \cite{Pan96}&$\!1.36~10^{-14}\!$&$\!4.32~10^{-15}\!$&$\!8.17~10^{-10}\!$&$\!-5.24~10^{-15}\!$&$\!-4.73~10^{-12}\!$&$\!-8.51~10^{-15}\!$\\
\hline
$^{130}Te$&&&&&&\\
&present&$\!7.83~10^{-13}\!$&$\!1.97~10^{-12}\!$&$\!5.66~10^{-8}\!$&$\!-5.19~10^{-13}\!$&$\!1.75~10^{-10}\!$&$\!-2.34~10^{-12}\!$\\
&ref. \cite{Mut89}&$\!5.34~10^{-13}\!$&$\!1.05~10^{-12}\!$&$\!2.25~10^{-8}\!$&$\!-2.17~10^{-13}\!$&$\!9.10~10^{-11}\!$&$\!-4.13~10^{-13}\!$\\
&ref. \cite{Pan96}&$\!3.02~10^{-13}\!$&$\!7.44~10^{-13}\!$&$\!1.61~10^{-8}\!$&$\!-2.28~10^{-13}\!$&$\!-6.24~10^{-11}\!$&$\!-1.49~10^{-12}\!$\\
\hline\end{tabular} \end{center}
\end{table}

The nuclei $^{76}Ge$, $^{82}Se$, $^{100}Mo$, $^{128}Te$ and $^{130}Te$ have
been evaluated within an eleven dimensional model space including all 
single particle orbitals of oscillator shells $3\hbar\w$ and $4\hbar\w$ plus
the $0h_{9/2}$ and $0h_{11/2}$ orbitals from the $5\hbar\w$ oscillator shell.
In the case of $^{48}Ca$ we work in a seven dimensional model space
including all the orbitals in the major shells $2\hbar\w$ and $3\hbar\w$.  
Here, the experimental single-particle energies have been used for the orbitals
$1p_{1/2}$, $0f_{5/2}$, $1p_{3/2}$, $0f_{7/2}$, $1s_{1/2}$ and $0d_{3/2}$,
while for the remaining orbitals a single-particle energy spacing of
$\hbar\w=41~A^{-1/3}$ MeV has been assumed.
Finally, both the $T=1$ and $T=0$
proton-neutron interaction strengths in the particle-particle
channel have been set by following the recipe introduced in ref. \cite{Hir90}.

Our results for nuclear matrix elements are compared in Table \ref{tab5}
with those obtained by MBK and PSVF.
In both works configuration spaces similar to ours were employed, and
the FNS effect was included in the way we have done it (see eq. \rf{23}).
 Yet there are two differences that could in principle be important:
i) instead of the  $\delta$-force, they have used the G-matrix (derived from the nucleon-nucleon
potential) as the residual interaction, and
ii) their correlation function is not that given by eq. \rf{22}.
In spite of these dissimilarities, our results
concord  surprisingly well with those obtained by MBK, except for
$M_{\sss T}$ and $M_{\sss P}$.
\footnote{Except the tensor moment $M_{\sss T}$,
the MBK matrix elements agree remarkably well with those
obtained by Tomoda and Faessler \cite{Tom87}. These authors have not
evaluated $M_{\sss T}$, since a negligible small value for it was obtained
previously
\cite{Tom86}
in a projected mean-field approach.}
The major difference is found in
$^{100}Mo$, but we know that this is a "difficult" nucleus from the
nuclear structure point of view, because of the collapse  of the QRPA
in the physical region of the particle-particle $T=0$ strength.
Moreover this effect is amplified by the SRC.
The agreement with
the PSVF calculation is only good in the case of the Gamow-Teller moments.
Note that the last can also be say for the concordance
between the MBK and PSVF results.

The coefficients $C_{ij}$, defined in eq. \rf{3} and evaluated with the matrix elements
given in Table \ref{tab5}, are compared in Table \ref{tab6}.
Kinematical factors from ref. \cite{Doi85}
have been used in our calculations. Obviously, all the above mentioned differences
between the matrix elements are reflected on the calculated
$C_{ij}$ values. However, the spread between the entries in the same row
in Table \ref{tab6} is smaller than the spread of the values in Table
\ref{tab5}. This is because the effect of the matrix elements
$M_{\sss T}$ and $M_{\sss P}$ is comparatively small.

Finally, Table \ref{tab7} gives the constraints on the Majorana neutrino mass and
the right-handed coupling constants, deduced from the most recent
experimental bounds for
the $\b\b_{0\n}$ half-lives, and the present evaluation of the nuclear matrix
elements. It should be kept in mind that in doing so we have used
the bare value $g_{\sss A}=1.254$ for the axial-vector coupling constant,
and that the upper limits for the lepton violating terms shown in Table
\ref{tab7} do not simply scale as $g_{\sss A}^4$.
\begin{table}[h]
\begin{center}
\caption {Experimental half-lives for the neutrinoless double beta
decay and upper limits on the Majorana neutrino mass $\ex{m_\n}$, and the
right-handed current coupling strengths $\ex{\l}$ and $\ex{\eta}$.}
\label{tab7}
\bigskip
\begin{tabular}{ccccc}
\hline
&&&&\\
Nucleus&$T_{0\nu}(exp)~[yr]$&$|\ex{m_\n}|~[eV]$&$|\ex{\l}|$&$|\ex{\eta}|$\\
&&&&\\
\hline\hline
$^{48}Ca$&$\!>~1.1~10^{22}~^{a)}\!$&$\!<~15\!$&$\!<~1.1~10^{-5}\!$&$\!<~7.9~10^{-8}\!$\\
$^{76}Ge$&$\!>~1.2~10^{25}~^{b)}\!$&$\!<~0.51\!$&$\!<~8.1~10^{-7}\!$&$\!<~3.2~10^{-9}\!$\\
$^{82}Se$&$\!>~2.7~10^{22}~^{a,b)}\!$&$\!<~5.3\!$&$\!<~5.7~10^{-6}\!$&$\!<~3.2~10^{-8}\!$\\
$^{100}Mo$&$\!>~5.2~10^{22}~^{a,b,c)}\!$&$\!<~3.4\!$&$\!<~3.2~10^{-6}\!$&$\!<~1.7~10^{-8}\!$\\
$^{128}Te$&$\!>~7.7~10^{24}~^{d)}\!$&$\!<~0.97\!$&$\!<~3.4~10^{-6}\!$&$\!<~6.4~10^{-9}\!$\\
$^{130}Te$&$\!>~8.2~10^{21}~^{e)}\!$&$\!<~6.4\!$&$\!<~7.9~10^{-6}\!$&$\!<~4.6~10^{-8}\!$\\
\hline\end{tabular} \end{center}
$^{a)}$ (laboratory data) ref. \cite{Mor96}

$^{b)}$ (laboratory data) ref. \cite{Kla98}

$^{c)}$ (laboratory data) ref. \cite{Eji96}

$^{d)}$ (geochemical data) ref. \cite{Ber92}

$^{e)}$ (laboratory data) ref. \cite{Ale94}
\end{table}
\newpage
\section {Concluding remarks}

Nuclear moments for the neutrinoless double beta decay have been evaluated
numerically for several nuclei, using the formalism that we have
recently developed. Simple analytic expressions for the $\b\b$ decay
of $^{48}Ca$, that follow from this formalism in the single mode model,
are also presented. The results shown in Table \ref{tab2} are useful,
not only for testing the full numerical calculations, but also for checking
the consistency  with other formalisms \cite{Hax84,Doi85,Ver90,Suh90}.
In fact, it would be highly desirable to find out whether these formalisms
lead to numbers shown in Table \ref{tab2}. This would be a simple and
definite test for all nuclear matrix element except for the $M_{\sss P}$.
However, even in this case, a simple model can be framed for confronting different
formalisms with each other.

The present work differs from similar QRPA studies, not only in the
$\b\b_{0\n}$ formalism, but also in the residual interaction. Namely, we have
used a simple $\delta$-force, instead of the G-matrix that is currently employed.
The fact that our results are equivalent to those obtained by MBK and PSVF
clearly shows that the $\b\b_{0\n}$ half-lives are not very sensitive to
details of the nuclear force. In other words, the so called "realistic
interactions" are in no way the panacea for the nuclear structure evaluation
of the $\b\b$ processes, as it has been proclaimed by some authors for a
long time.
The reason for that is the crucial role played by the restoration of the isospin
and SU(4) symmetries, produced by the residual interaction, in tailoring the
Fermi and Gamow-Teller transitions strengths, respectively \cite{Krm94}.
This restoration mechanism is not limited to the RPA-like models
\cite{Krm94,Eng70,Lee71,Lan80}, but also occurs in the shell model calculation.
(One simple example has been discussed in Sec. 3.) We are convinced that the nuclear
structure issue, involved in $\b\b$-decays will still keep us busy for a
long time, and that definitively it cannot be solved by a simple minded
employment
of "good" residual interactions. Neither the renormalized nor self consistent RPA
methods are able to fix this problem \cite{Krm98a}.

An alternative technique for examining the $\beta \beta$ transitions could be
the relativistic RPA, which has recently been successfully applied to
the description of the isobaric analogue and Gamow-Teller resonances in closed
shell nuclei \cite{Con98}.
We are planning to analyze the consequences of such an approach.

\end{document}